\documentclass[preprint,pre,aps,preprintnumbers,amsmath,amssymb,superscriptaddress]{revtex4}

\usepackage{amssymb,amsfonts,amsmath}
\DeclareMathOperator{\arctanh}{arctanh}
\usepackage{graphicx}
\usepackage{longtable}
\usepackage{color} 
\usepackage{setspace}
\usepackage{url}

\begin{document}
{\singlespace
\title{Failure dynamics of the global risk network}

 \author{Boleslaw K. Szymanski  \footnote{Corresponding author: szymab@rpi.edu} }
\affiliation{Social and Cognitive Networks Academic Research Center, Rensselaer Polytechnic Institute, Troy NY 12180} 
\affiliation{Dept. of Computer Science, RPI, 110 8th Street, Troy, NY 12180}

\author{ Xin Lin}
\affiliation{Social and Cognitive Networks Academic Research Center, Rensselaer Polytechnic Institute, Troy NY 12180} 
\affiliation{Dept. of Computer Science, RPI, 110 8th Street, Troy, NY 12180}

\author{Andrea Asztalos  \footnote{Current address: NCBI, NLM, NIH Bethesda, Maryland USA}}
\affiliation{Social and Cognitive Networks Academic Research Center, Rensselaer Polytechnic Institute, Troy NY 12180} 
\affiliation{Dept. of Computer Science, RPI, 110 8th Street, Troy, NY 12180}
\affiliation{Dept. of Physics, Applied Physics and Astronomy, RPI, 110 8th Street, Troy, NY 12180}

\author{Sameet Sreenivasan}
\affiliation{Social and Cognitive Networks Academic Research Center, Rensselaer Polytechnic Institute, Troy NY 12180} 
\affiliation{Dept. of Computer Science, RPI, 110 8th Street, Troy, NY 12180}
\affiliation{Dept. of Physics, Applied Physics and Astronomy, RPI, 110 8th Street, Troy, NY 12180}

\begin{abstract} 
Risks threatening modern societies form an intricately interconnected network that often underlies crisis situations. Yet, little is known about how risk 
materializations in distinct domains influence each other. Here we present an approach in which expert assessments of risks likelihoods and influence underlie 
a quantitative model of the global risk network dynamics. The modeled risks range from environmental to economic and technological and include difficult 
to quantify risks, such as geo-political or social. Using the maximum likelihood estimation, we find the optimal model parameters and demonstrate that the 
model including network effects significantly outperforms the others, uncovering full value of the expert collected data. We analyze the model dynamics and 
study its resilience and stability. Our findings include such risk properties as contagion potential, persistence, roles in cascades of failures and the
identity of risks most detrimental to system stability. The model provides quantitative means for measuring the adverse effects of risk interdependence and 
the materialization of risks in the network.

\end{abstract}

\maketitle

Modern society relies heavily on the robust functioning of systems that are intricately networked with one another, either in an explicit or an implicit 
manner. While increasing the interconnectivity between infrastructure systems can result in a higher efficiency of service, it also makes the constituent 
systems vulnerable as a whole to cascading failures. Such cascades of failures have been studied generally in model networks 
\cite{Watts2002, MotterLai2002,Buldyrev2010,Asztalos2012, Brummitt2012,Roukny2013} and specifically in the context of engineered systems such as the 
power-grid \cite{Dobson2004}, the internet \cite{Oppenheimer2003} and transportation and infrastructure systems \cite{Atalay2011}, in the context of 
financial institutions \cite{Gai2010,Haldane2011,Battiston2012,Battiston2013,Huang2013}, and within ecological systems \cite{Schmitz2000}. However, 
in addition to the risk of cascading failures being present within a particular domain (e.g., the network of financial institutions), there are also risks 
arising because of the coupling between systems in diverse domains \cite{Vespignani2010, Helbing2013}. Indeed, the primary thesis behind many societal 
collapses in the history of mankind is that of a cascade of diverse risks being materialized \cite{Diamond2004}. Examples of such cross-domain failure 
cascades include the collapse of the society on Easter Island stemming from deforestation that led to agricultural and economic instabilities and civil 
unrest, and the demise of the populations of the Pitcairn and Henderson Islands caused by an environmental catastrophe of their common but geographically 
distant trading partner of Mangareva. The acceleration of technological advances over the last two centuries and the virtual dissolution of geographical 
borders resulted in tightening of the coupling between risks in diverse domains and across geographically distant systems. The recent economic crisis and 
its widespread effects across the globe have demonstrated this all too clearly. Hence, there is an urgent need to quantify the dynamics of large scale 
risk materialization lurking within this globally interconnected tapestry. Moreover, enriching our understanding of systems formed by diverse, 
interconnected sub-systems spanning environmental, social, and infrastructural domains is an important component of the scientific study of the physical world.

As a qualitative means to this end, the World Economic Forum (WEF) publishes each year a report defining the network of global risks \cite{wefreportlink}. 
The report published in 2013 contains data on the likelihood of materialization of global risks, influence of risk materialization on other risks, and 
the potential impact that materialization of each risk has on the global economy. This data was prepared by over 1000 experts from government, industry, 
and academia. 

The risks defined in the report are classified into five categories: economic, environmental, geopolitical, societal, and technological. This global risk 
network dataset thus provides an experts' perspective on the threat of different risk materializations and their influence on other risks, both of which 
are often intangible to a non-expert.

The collection of evaluations made by large groups of experts is often termed crowdsourcing. Its value has been well documented through the rise of online 
encyclopedias and question-answer sites, and the use of tagging or classification by groups of experts for recommendation-based services such as Pandora 
and Netflix. Consequently here, we use the WEF global risk dataset as a starting point to performing a quantitative study that can generate actionable 
insights. Specifically, we propose a model for the materialization of risks on the network, in which internal (self-materialization) and external (contagion) 
risk materializations are separated using historical data. This allows us to estimate the parameters associated with the network, and thereby to define the 
dynamics of global risk network. Furthermore, we show that a model incorporating the influence of a risk's network neighborhood on its own materialization 
matches past data on risk materialization better than a model which is oblivious to network effects. We analyze the model dynamics and study its resilience, 
stability, and risks contagion potential, persistence, and roles in cascades of failures, identifying risks most detrimental to system stability. 
The model provides quantitative means for measuring the adverse effects of risk interdependence and the materialization of risks in the network. Such 
quantitative insights can, in turn, form a sound basis for specific recommendations from domain experts on how to decouple the severely harmful risks from 
the network, or to reduce their materialization probabilities. 

\section*{Results}

\subsection*{Dataset}
We utilize two datasets for this study, the first of which is a part of the WEF Report on Global Risks 
released in 2013 and available at \cite{wefdatalink}. The report defines $N=50$ global risks (see \cite{wefreportlink} for their detailed definitions)
that are systemic, that is they ``breakdown 
in an entire system, as opposed to breakdowns in individual parts and components'' \cite{Kaufman2003}. As pointed our in \cite{Helbing2013}, each of 
these risks is a network itself.  Together they represent a network of networks that are prone to catastrophic cascades of failures 
\cite{Buldyrev2010}. These failures are binary in nature, since they either manifest themselves or not (even though the size, range and economic 
impact of each manifestation may vary from one manifestation to another).
The WEF Report on Global Risks \cite{wefreportlink} includes expert assessments of the 
likelihood of each risk materialization in the next decade and of the influence of each risk materialization on other risks. Each expert defines these likelihoods 
by numerical value ranging from  $1$ (lowest likelihood) to $5$ (highest likelihood) using integers and  
mid-points between integers within this range. 
Thus, although not explicitly stated, the assessment of risk materialization 
likelihoods by the experts was done on a quantitative scale with a resolution 
corresponding to probability increments of $1/8$.
 
The averages and standard deviations of the expert likelihood values for each risk are also
provided in the dataset. Thanks to high quality and diversity of the experts participating in this crowdsourced assessment, the margins 
of error are small. This is demonstrated in Table 3, Appendix 2, page 66 of \cite{wefreportlink} that lists the average likelihood and impact scores 
and shows that their margins of error (based on a 95\% confidence level) vary from 1\% to 2\% of the average score values. 
The detailed analysis of the data collected in the dataset is provided in Appendices 1 and 2 of \cite{wefreportlink}.

We denote the average of assessments for risk $i$ by $L_i$.  The dataset provides a list of pairs of risks, aggregated from 
expert selections of up to ten most influential pairs, such that both risks in each pair are perceived as having influence on each other. 
The weight of influence for each pair of the selected nodes is also listed and it represents the number of experts that chose the corresponding pair
as influential. In the risk network, the selected pairs are represented as connected. We selected the model which uses just binary value 
representing whether two nodes are connected or not since this information yielded the same maximum expectation when optimized using historical data 
as the models that use the weights of connections. Accordingly, we designate a 
binary variable $b_{ij}$ to each pair of risks $(i,j)$ that is $1$ if the pair appears at least once in the concatenated lists of the experts, and $0$ otherwise. 
Thus, $b_{ij}$ captures some overall {\it association} between risks $i$ and $j$. By definition $b_{ij} = b_{ji}$. There are $515$ bidirectional edges defined 
that way, thus, the average degree of each node is relatively high, $20.6$ edges per node. 

The entire global risk graph can be modeled as a Stochastic Block Model \cite{Holland1983} 
with probability $p_g$ of the edge existing between two risks in the same group $g$ (this probability differs from 
group to group) while connections between nodes from two different risk groups, $(g1,g2)$, are drawn with unique probability $p_{g1,g2}$. 
The values of these probabilities for the WEF global risk graph are shown in Fig.~\ref{RGdistr} and~\ref{networkviz}. 
The adjacency matrix of this graph is denoted $A$, and its binary element $a_{i,j}$ is $1$ if and only if edge $(i,j)$ exists.   
\begin{figure*}[!htb]
\begin{center}
\includegraphics[height = 5in,angle=90]{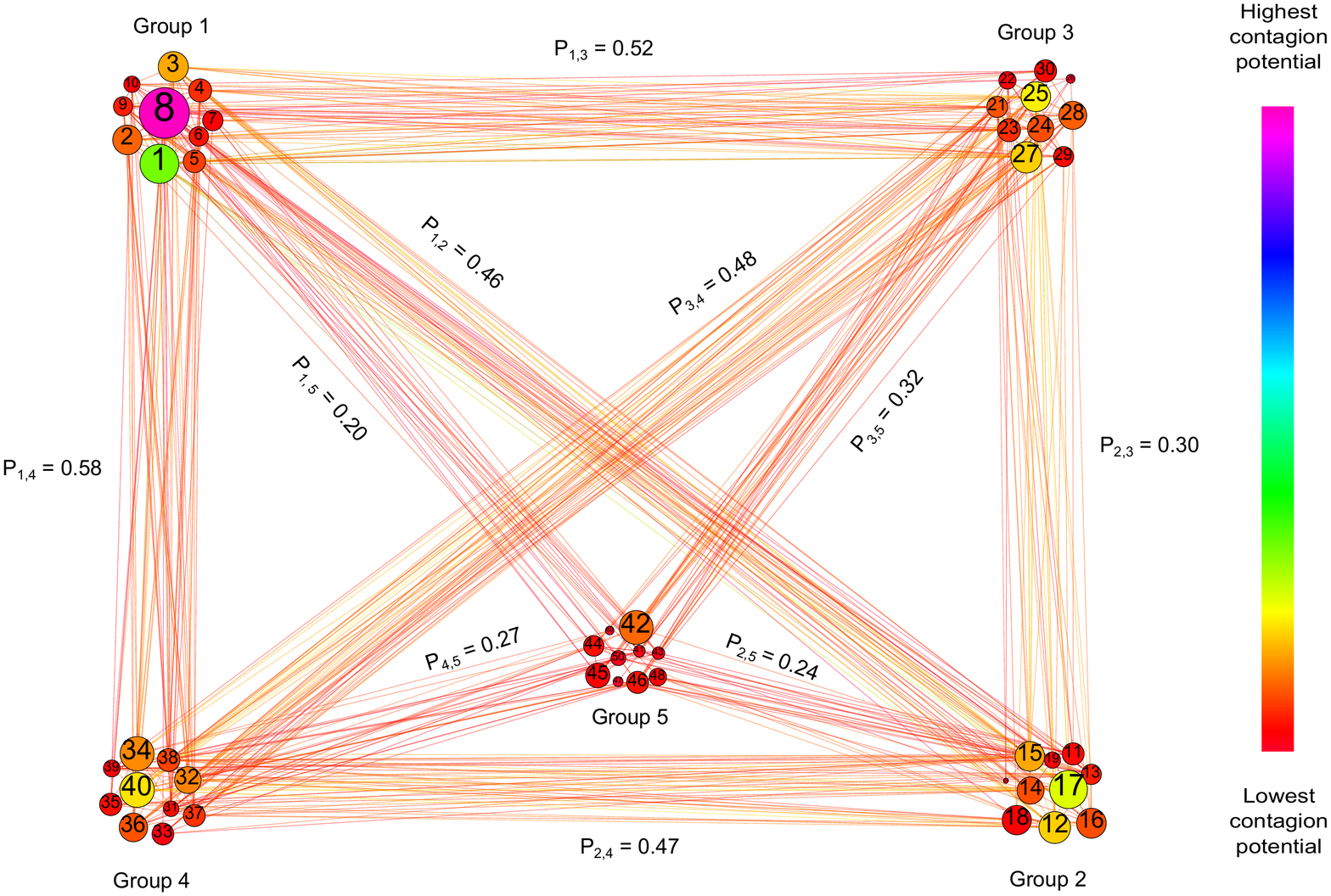}
\end{center}
\vspace{-0.4in}
\caption{{\bf Connectivity of Risk Groups:} Visualization of inter-group connectivity where node's color corresponds to
its total connectivity and the number of lines connecting the groups indicate strength of the inter-group connectivity. Groups $1$ (economic risks), $2$ (environmental risks), 
and $3$ (geopolitical risks) are the best connected, so risks from these groups dominate the list of most persistent nodes. The remaining 
two groups:
$4$ (societal risks) and $5$ (technological risks), have fewer connections to other groups. The inter-edges are labeled 
with probability of inter-group connections (intra-group edges are shown in Fig. \ref{networkviz}). 
}
\label{RGdistr}
\end{figure*}
We express the probabilities of risk materialization in terms of the $L_i$s and $b_{ij}$s obtained from the WEF dataset and the parameters of our model.

In order to optimize parameters for our model, we require a time series of risk materialization events over a contiguous period. We collected data on the 
materialization of each of the $50$ risks over the period $2000-2012$ by systematically surveying news, magazine and academic articles and websites identified 
using targeted queries on Google, as well as Wikipedia entries pertinent to specific risk materialization events. For brevity, we refer to this dataset as the 
{\it historical} dataset. Overall, it provides $50\times 156=7,800$ data points for tuning the system parameters.
\begin{figure*}[!htb]
\begin{center}
\includegraphics[height = 5in,angle=90]{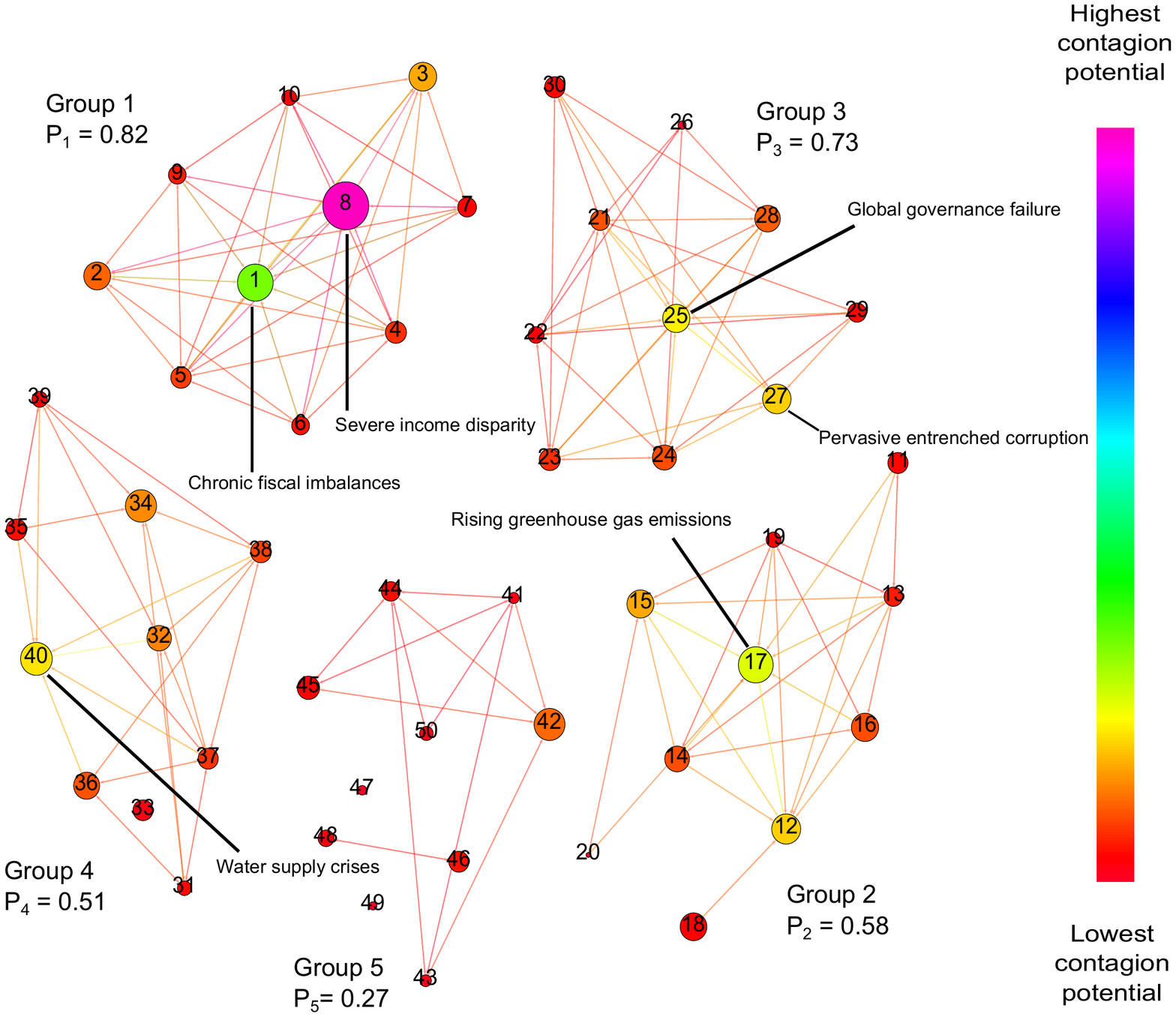}
\end{center}
\vspace{-0.4in}
\caption{{\bf Global risk network intra-group connectivity and nodes congestion potentials derived from optimal model parameters:}  
Each node is sized proportionally to its internal failure probability while node color corresponds to its total contagion potential. 
The number of edges in each group shows the intra-group connectivity. The nodes with the highest 
congestion potential are identified by name and include risks $8$ ``Severe income disparity'', $17$ 
``Rising greenhouse gas emissions'', $1$ ``Chronic fiscal imbalances'', $40$ ``Water supply crises'', $25$ 
``Global governance failure'' and $27$ ``Pervasive entrenched corruption''.  
}
\label{networkviz}
\end{figure*}

\subsection*{Model}

The failure dynamics of the global risk network can be modeled using Alternating Renewal Processes (ARP) \cite{Cox1965}, which were initially used 
for engineered systems, but recently have been applied to science and economic applications \cite{Beichelt2006}. Nodes under 
ARP alternate between normal state and failure state, and the corresponding events of failure activation and recovery are responsible for the 
state transitions. Typically, these events are assumed to be triggered by the non-homogeneous Poisson Processes (i.e. Poisson Processes with time 
dependent intensities) \cite{Cox1965}. In our systems, use of the Poisson distribution is justified because each failure represents a systemic risk 
that can be triggered by many elementary events distributed all over the globe. Such triggering distorts any local temporary patterns of events (such as 
periodic weather related local disasters in some regions of the globe). Moreover, the time-dependence of intensity of risk activation is the results 
of influence that active risks  exert on passive risks connected to them. Accordingly, each risk at time $t$ is either in state $1$ (materialized 
or {\it active}) or state $0$ (not materialized or {\it inactive}). 

In a traditional ARP, there are two  directly observable processes, one of risk activation and the other of recovery from the active risk. However, 
in our model, we introduce two latent processes that together represent the risk activation process.  As explained later, we use the 
Expectation-Maximization Algorithm \cite{Dempster1977} to find model parameters that make it optimally match historical data.
Consequently, we assume that changes in the state of each risk result from events generated by three types of Poisson processes. 
First, for a risk $i$, given that it is in state $0$, its {\it spontaneous} or {\it internal} materialization is a Poisson process with intensity 
$\lambda^{\rm int}_i$. Similarly, given the risk in state $1$, its recovery from this state, and therefore transition to state $0$, is a Poisson process with 
intensity $\lambda^{\rm rec}_i$. Finally, given that risks $i$ is in state $0$, and $j$ in state $1$ are connected, the materialization of risk $i$ due to the 
{\it external} influence of risk $j$ is a Poisson process with intensity $\lambda^{\rm ext}_{ji}$. We assume that each of these processes is independent of each 
other. We also evaluated models in which recovery is represented by two latent processes, one of internal recovery 
and the other of recovery induced by either the connected passive or active risks. In both cases, the optimal intensity of the externally 
induced recovery was $0$. Thus, the simpler model with just internal recovery is used as it yields the same results as the more complex models
using latent processes for recovery.

The model is similar to a model of house fires in a city, where some houses burn alone from a self-started fire, but others are ignited by the 
burning neighboring houses. Yet, the recovery, that is rebuilding of a burnt house, is independent of the state of its neighboring houses. 

Denoting by $s_i(t)$ the probability at time $t$ that the state of a risk $i$ at that time is $1$, we can express the expected number of risks 
materialized at time $t$ as the sum of all $s_i(t)$'s, each of which is defined by the following Ordinary Differential Equation (ODE):  
\begin{equation}
 \frac{ds_i(t)}{dt}=\lambda^{\rm int}_i(1-s_i(t))-\lambda^{\rm rec}_is_i(t)+\lambda^{\rm ext}_i(1-s_i(t))\sum_{j=1,j\neq i}^Na_{i,j}s_j(t)
 \label{ODE}
 \end{equation}
Checking stability, we conclude that this system of non-linear ODEs has only one unique stable point in the feasible range $0\leq s_i(t)\leq 1$, 
which can easily be found numerically. Moreover, this system of ODEs for a fully connected graph when the intensities 
$\lambda_s, \lambda_r, \lambda_e$ of the three Poisson processes are independent of the node on which they operates, and for all nodes starting 
in the same initial condition $s(0)$ has the analytic solution of the form 
$s_i(t)=\frac{a+b*\tanh(b*t/2+\arctanh((2\lambda_E*s(0)+a)/b))}{2\lambda_E}$, where $\lambda_E=(n-1)\lambda_e$, $a=\lambda_s+\lambda_r-\lambda_E$, 
and $b=\sqrt{a^2-4\lambda_s\lambda_E}$. This solution tends asymptotically to $\frac{2\lambda_s}{\lambda_s+\lambda_r
-\lambda_E-\sqrt{(\lambda_s+\lambda_r-\lambda_E)^2-4\lambda_s\lambda_E}}$. 

For nearly all events that we consider here, it is difficult to assign precise starting and ending times for their periods of materialization. Thus, 
it is more proper to consider a Bernoulli process in which the time unit (and also time step of the model evolution) is one calendar month (we 
ignore the minor numerical imprecision arising from the fact that calendar months have different numbers of days). Consequently, all events starting in the 
same month are considered to be starting simultaneously. Hence, at each time step $t$, each risk $i$ is associated with a binary state variable $S_i(t) \in \{0,1\}$. 
The state of the entire set of risks at time $t$ can therefore be represented by a state vector $\vec{S}(t)$.  
Thus, the dynamics progresses by assuming that at
each time step $t>0$:
\begin{enumerate}
\item a risk $i$ that was inactive at time $t-1$ materializes internally with probability 
$p^{\rm int}_i=1-e^{-\lambda^{\rm int}_i}$.
\item a risk $j$ that was active at time $t-1$ causes a connected to it risk $i$ that was inactive at time $t-1$
to materialize with probability $p^{\rm ext}_{ji}= 1-e^{-\lambda^{\rm ext}_{ji}}$.
\item a risk $i$  that was active at time $t-1$ continues its materialization with probability $p^{\rm cont}_i=e^{\lambda^{\rm rec}_i}=1-p^{\rm rec}_i$.
\end{enumerate}
It is easy to show that for real time $t$ measured in finite time units (months in our case), the Poisson process assumption for events results in a probability 
$1-e^{-\lambda \lceil{t}\rceil}$ of an event happening in at most $\lceil{t}\rceil$ time units which is identical to the assumed Bernoulli process. The advantage 
of the latter process is that in each step the probability of event is known, simplifying maximum likelihood evaluation of the model parameters.
Finally, the dynamics described above imply that the state of the system at time $t$ depends only on its state at time $t-1$, 
and therefore the evolution of the state vector  $\vec{S}(t)$ is Markovian. 

Given the probabilities of internal materialization, external influence and internal continuation, that is just 1 minus the probability of recovery, 
the probability of a transition in a risk's state between consecutive time steps can be written in terms of these probabilities:
\begin{equation}
\begin{array}{l}
\displaystyle \mathcal{P}_i(t)^ {0 \rightarrow 1} = 1 - e^{-\lambda^{\rm int}_i-\sum_{j \in A(t-1)}\lambda^{\rm ext}_{ji}}\\
\displaystyle \mathcal{P}_i(t)^{0 \rightarrow 0} = 1 - \mathcal{P}_i(t)^{0 \rightarrow 1} \\
\displaystyle \mathcal{P}_i(t)^ {1 \rightarrow 0}  = 1 - e^{-\lambda^{\rm rec}_i}\\
\displaystyle  \mathcal{P}_i(t)^ {1 \rightarrow 1} = 1 - \mathcal{P}_i(t)^ {1 \rightarrow 0}
\label{transitions}
\end{array} 
\end{equation}

where $A(t)$ represents the risks that are active at time $t$, and $\mathcal{P}_i(t)^{x \rightarrow y}$ is the 
probability that risk $i$ transitions between time $t-1$ and $t$ from state $x$ to state $y$, or in other words 
$S_i(t-1) = x$ and $S_i(t) = y$. 

The mapping of the Poisson process intensities into the expert assessments is described in the Methods section. Each of the probabilities of Bernoulli processes
is mapped onto the probability obtained from expert assessment of likelihood of risk failure by single-parameter formula. We find the values of the model 
parameters that optimize the model match with the historical data, while we use expert assessments to individualize probabilities of Bernoulli processes for
each risk. In essence, the expert assessments are defining those probabilities for each risk in relations to probabilities for other risks, while model 
parameters map performance of all risks onto historical data. By distinguishing between internal and external materialization factors, the mapping of parameters 
onto historical data enables us also to decompose risk materializations into these two categories. Once the mapping is done, the model is complete and can be 
used to evaluate global risk dynamics. 

From several alternative models discussed in Methods section, we discuss below the best performing network model which uses all three parameters, and the 
disconnected model which sets the value of probability of influence of a risk materialization on any other risk to zero, effectively isolating risk 
materializations from each other. 

\subsection*{Contagion potentials of risks in the network model}
\label{contagious-potential-section}
Here, we investigate the relative importance of different risks.
First, in analogy with epidemic studies, we calculate the {\it contagion potential} of individual risks, i.e., the mean number of materializations that 
a risk induces given that it alone has materialized. For risk $i$, the exact expression for this quantity is:

\begin{equation}
 C_i = \sum_{j=1, j \neq i}^N \frac{p^{\rm con}_i p_{ij}^{\rm ext}} {1-p_{i}^{\rm con}+p_{i}^{\rm con} p^{\rm ext}_{ij}}
 \label{contagionpot}
 \end{equation}

where $N$ refers to the total number of risks. This expression assumes that risks other than $i$ can only be activated through the influence of risk $i$ 
and not internally.

Figure~\ref{networkviz} shows a visualization of the network capturing the contagion potentials as well as the internal failure probabilities in the 
network model (the mapping of node indices to the risks is provided in Table~\ref{risk-mapping}). As illustrated, the internal failure probability 
does not strictly show a positive correlation with contagion potential. Hence, a frequently materializing risk does not necessarily inflict the most 
harm to the system as a result of its influence on other risks. For example, although risk $42$ - ``Cyber attacks" has a relatively high probability 
of internal materialization, its contagion potential is low. In contrast, risk $25$ - ``Global governance failure" has both a high probability of 
internal materialization and a high contagion potential. However, most striking is the fact that risk $8$ - ``Severe income disparity'' has both the 
highest internal materialization probability and the highest contagion potential.  This is particularly notable in the light of recent claims that 
income disparity in the United States is the highest in over eight decades \cite{Saez2013}. 

The five risks with the highest contagion potentials are: $8$ - ``Severe income disparity",  $1$ - ``Chronic fiscal imbalances", 
$17$ - ``Rising greenhouse gas emissions'', $40$ - ``Water supply crises", and $12$ - ``Failure of climate change adaptation''. When ranked purely by raw 
likelihood values $L_i$ (or equivalently by the internal failure probabilities $p_i^{\rm int}$), the only change is on the fifth position, where risk $12$ is
replaces by risk $34$ - ``Mismanagement of population ageing" moves up from eleventh position to fifth.



\begin{figure*}[!htb]
\vspace{-0.5in}
\begin{center}
\includegraphics[width = 4.5in]{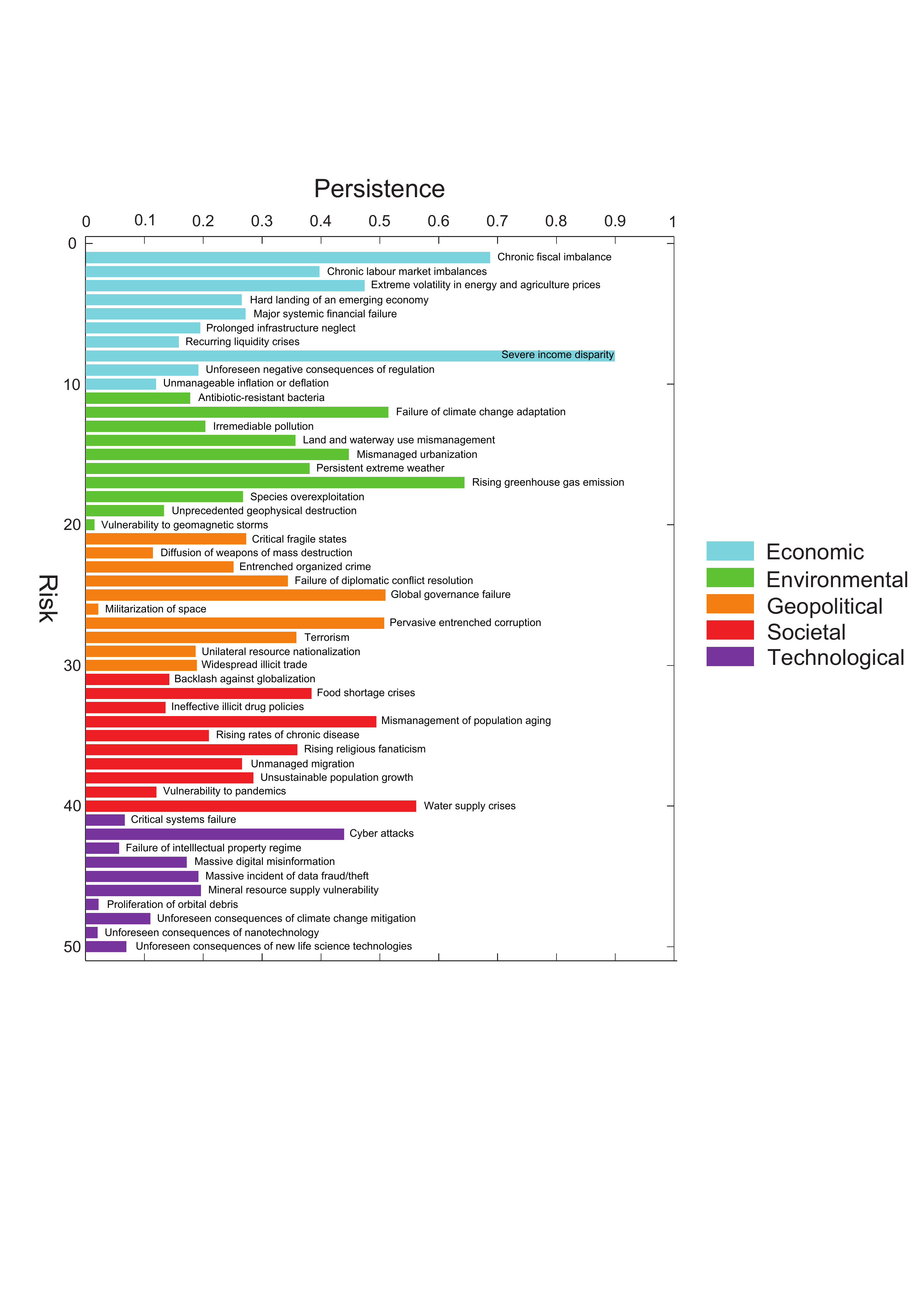}
\end{center}
\vspace{-1.75in}
\caption{{\bf Persistence of risks in model simulation:} The bar graph shows the overall fraction over $10^6$ simulated experiments, each 
consisting of 
$2200$ time steps of which the initial $200$ are ignored, that a given risk was active. The top five most persistent risks are $8$ - 
``Severe income disparity'', $1$ - ``Chronic fiscal imbalances'', $17$ - ``Rising greenhouse gas emissions", $40$ - ``Water supply crises",
and $12$ - 
``Failure of climate change adaptation". Not surprisingly this is the same lists as the list of the most contagious risks identified 
in the text because symmetric influence relationship makes them also most vulnerable to contagion. 
}
\label{persistencenormal}
\end{figure*}
\pagebreak
\begin{longtable}{|p{.03\textwidth}|p{.46\textwidth}|p{.03\textwidth}|p{.46\textwidth}|}
\hline
ID & Risk & ID & Risk\\
\hline
1 & Chronic fiscal imbalances & 2 & Chronic labour market imbalances \\
3 & Extreme volatility in energy and agriculture prices & 4 & Hard landing of an emerging economy \\ 
5 & Major systemic financial failure & 6 & Prolonged infrastructure neglect\\
7 & Recurring liquidity crises & 8 & Severe income disparity \\
9 & Unforeseen negative consequences of regulation & 10 & Unmanageable inflation or deflation\\
11 & Antibiotic-resistant bacteria & 12 & Failure of climate change adaptation \\
13 & Irremediable pollution & 14 & Land and waterway use mismanagement \\
15 & Mismanaged urbanization & 16 & Persistent extreme weather \\
17 & Rising greenhouse gas emissions & 18 & Species overexploitation \\
19 & Unprecedented geophysical destruction& 20 & Vulnerability to geomagnetic storms \\
21 & Critical fragile states & 22 & Diffusion of weapons of mass destruction \\ 
23 & Entrenched organized crime & 24 & Failure of diplomatic conflict resolution \\
25 & Global governance failure & 26 & Militarization of space \\
27 & Pervasive entrenched corruption & 28 & Terrorism \\
29 & Unilateral resource nationalization & 30 & Widespread illicit trade \\
31 & Backlash against globalization & 32 & Food shortage crises \\
33 & Ineffective illicit drug policies & 34 & Mismanagement of population aging \\
35 & Rising rates of chronic disease & 36 & Rising religious fanaticism \\
37 & Unmanaged migration & 38 & Unsustainable population growth \\
39 & Vulnerability to pandemics & 40 & Water supply crises \\
41 & Critical systems failure & 42 & Cyber attacks \\
43 & Failure of intellectual property regime & 44 & Massive digital misinformation \\
45 & Massive incident of data fraud/theft & 46 & Mineral resource supply vulnerability \\
47 & Proliferation of orbital debris & 48 & Unforeseen consequences of climate change mitigation \\
49 & Unforeseen consequences of nanotechnology & 50 & Unforseen consequences  of new life science technologies \\
\hline

\caption{\bf Mapping of indices to risks used throughout the paper}
\label{risk-mapping}
\end{longtable}


\subsection*{Network activity level and risk-persistence}
Next, we perform Monte-Carlo simulations of both the network model and the disconnected model. 
Figure ~\ref{persistencenormal} shows the fraction of time steps over $10^6$ simulations, each consisting $2200$ time steps, that a given risk was active 
(the initial transient consisting of $200$ steps was ignored). We call this fraction the {\it persistence of the risk}. Each simulation was initiated with 
the same active risks that are present in the first month of historical data (i.e. January 2000). The most persistent risk was $8$ active $90\%$  of the time, 
followed by risk $1$, active $68\%$ of the time, risk $17$, active $64 \%$ of the time, risk $40$, active $56$\% of the time, and risk $12$, active $51\%$ 
of the time. 

Another interesting aspect is the distribution of the number of active risks obtained in the simulation. The $10^{\rm th}$ percentile value of the number 
of active risks is below $8$, while the $90^{\rm th}$ percentile value of the number of active risks is over $19$, implying that about $80 \%$ of the time, 
the number of active risks will lie between these two values.

The steady state (long-time limit) activity levels indicate that the  {\it carrying capacity} of the global risk network at the present time is $27 \%$ 
of the size of the network, i.e., about $13.8$ risks are active all the time. The top seven risks observed to be active most frequently in simulations are 
$8, 1, 17, 40, 12, 25$, and $27$.  These seven risks contribute on average $4.3$ members to the total activity level at any month. This implies that other 
$43$ risks together contribute on average the remaining $9.4$ active risks, thus their activity level per risk is nearly three times lower than it is for the 
top seven risks.

\subsection*{Cascades due to single risk materializations}
We further study the effect of risk interconnectivity by investigating the survival probability of a failure cascade initiated by a particular risk's 
materialization. Specifically, we perform $10^6$ Monte-Carlo simulations of the model, each running for $50,000$ time steps, starting with a given single 
risk active and setting the internal failure probabilities of all risks to zero. Thus, all subsequent risk materializations (after the initial one) are 
caused purely by the {\it cascade} propagating within the network. Note that this is different from the true activity dynamics within the network discussed 
previously. These simulations are carried out to demonstrate the extent to which the connectivity between risks facilitates secondary activations. Shown in 
Figure~\ref{survivalprob} are the survival probabilities for cascades initiated by five highly contagious risks ranked in descending order of contagion 
potential. The linear nature of the curves on the linear-logarithmic scale indicates that survival probabilities decay exponentially with time. Despite 
that, even the cascade initiated by the {\it least} contagious risk in the displayed data, risk $40$ ``Water supply crisis", has a greater than $1$\% 
chance of continuing beyond $10,000$ months, i.e., over eight centuries. These long cascade lifetimes, even in the absence of internal failures, 
demonstrates the profound disadvantage of interconnectivity of global risks. 
\begin{figure*}[!htb]
\vspace{-1.75in}
\begin{center}
\includegraphics[width = 4.5in]{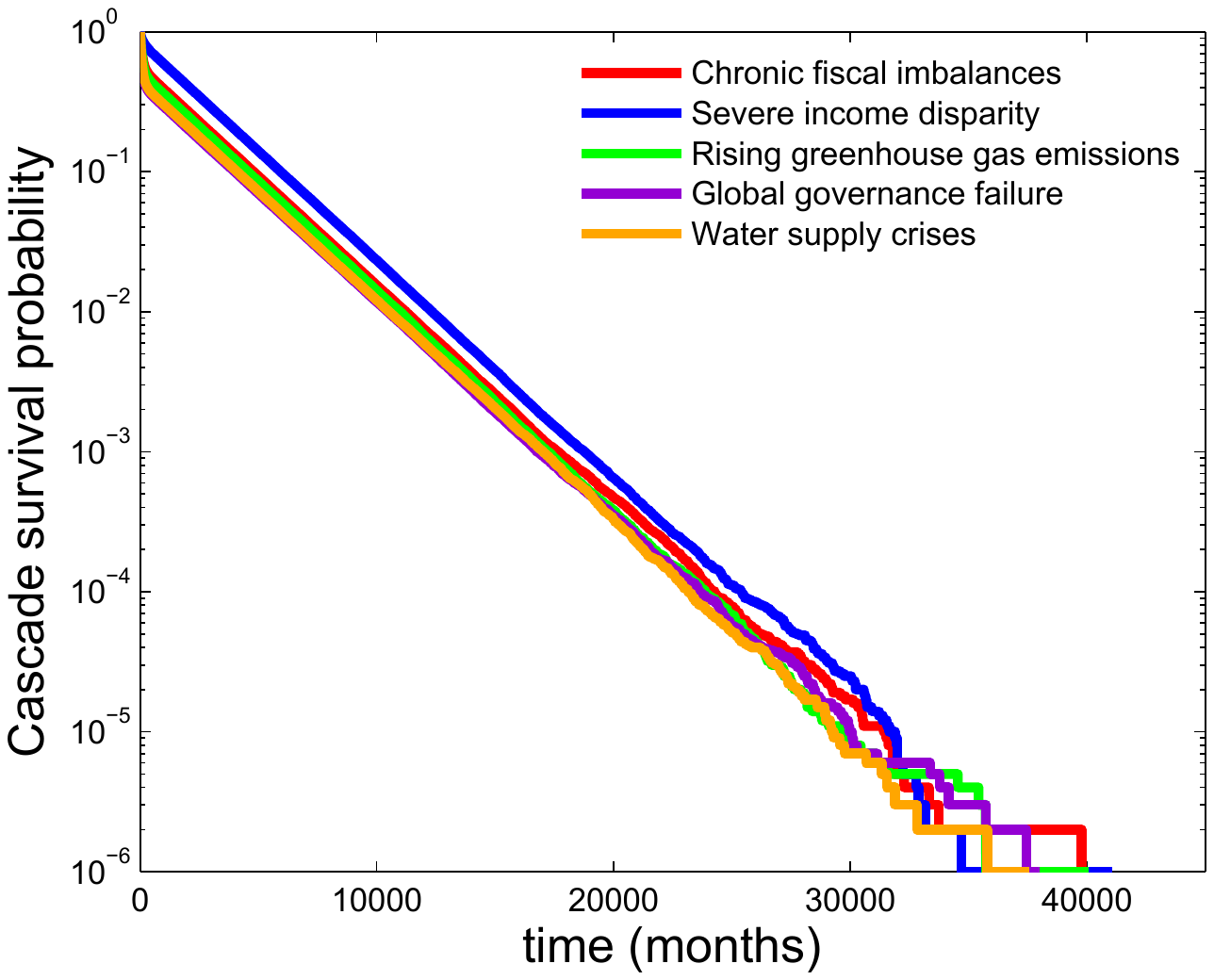}
\end{center}
\vspace{-2.25in}
\caption{{\bf Survival probability of a risk cascade initiated by a single failure:} Each curve shows the survival probability of a cascade initiated by 
a specific risk as a function of time. Five risks with high contagion potentials (listed in the text) were chosen as the respective cascade initiators, 
and the number of surviving cascade realizations among a total of $10^6$ realizations was computed for each chosen initiator. The straight line on the 
linear-logarithmic scale shows clear evidence of an exponentially decaying survival probability.
}
\label{survivalprob}
\end{figure*}
%
%

Next, we investigate which risks are predominantly responsible for the cascades persisting for such long time-scales. Figure~\ref{persistenceplot} shows 
the expected fraction of the lifetime of a cascade for which a particular risk is active, in ranked order. The top five highest active risks are $8$, 
active for $83$\% of the cascade lifetime, $1$, active for $53$\% of the lifetime, $17$, active for $46\%$ of the lifetime, $40$, active for $39$\% of 
the lifetime, and $12$, active for $35\%$ of the lifetime. Interestingly, the lists of top five most persistent risks observed in the cascades and seen 
in the full dynamics of activation (when all nodes undergo both internal and external activation Poisson processes) are identical.

We also compute the probability that the cascade resulting from the materialization of a given initiator risk would result in the materialization of 
a selected risk. Specifically, we consider the probability of materialization of the four risks, $8$, $1$, $17$ and $25$, observed to be among the top 
five risks most frequently active in simulations.  Risk $8$ is the initiator that yields the highest materialization probability for risks $1$, $17$, 
and $40$, while it itself materializes with highest probability when the initiator risk is $1$ and is followed by risks $17$, $25$, and $40$. The risks 
$8$ and $1$ materialized with highest probability for initiators $17$ and $25$.
\begin{figure*}[!htb]
\vspace{-0.4in}
\begin{center}
\includegraphics[width = 5in]{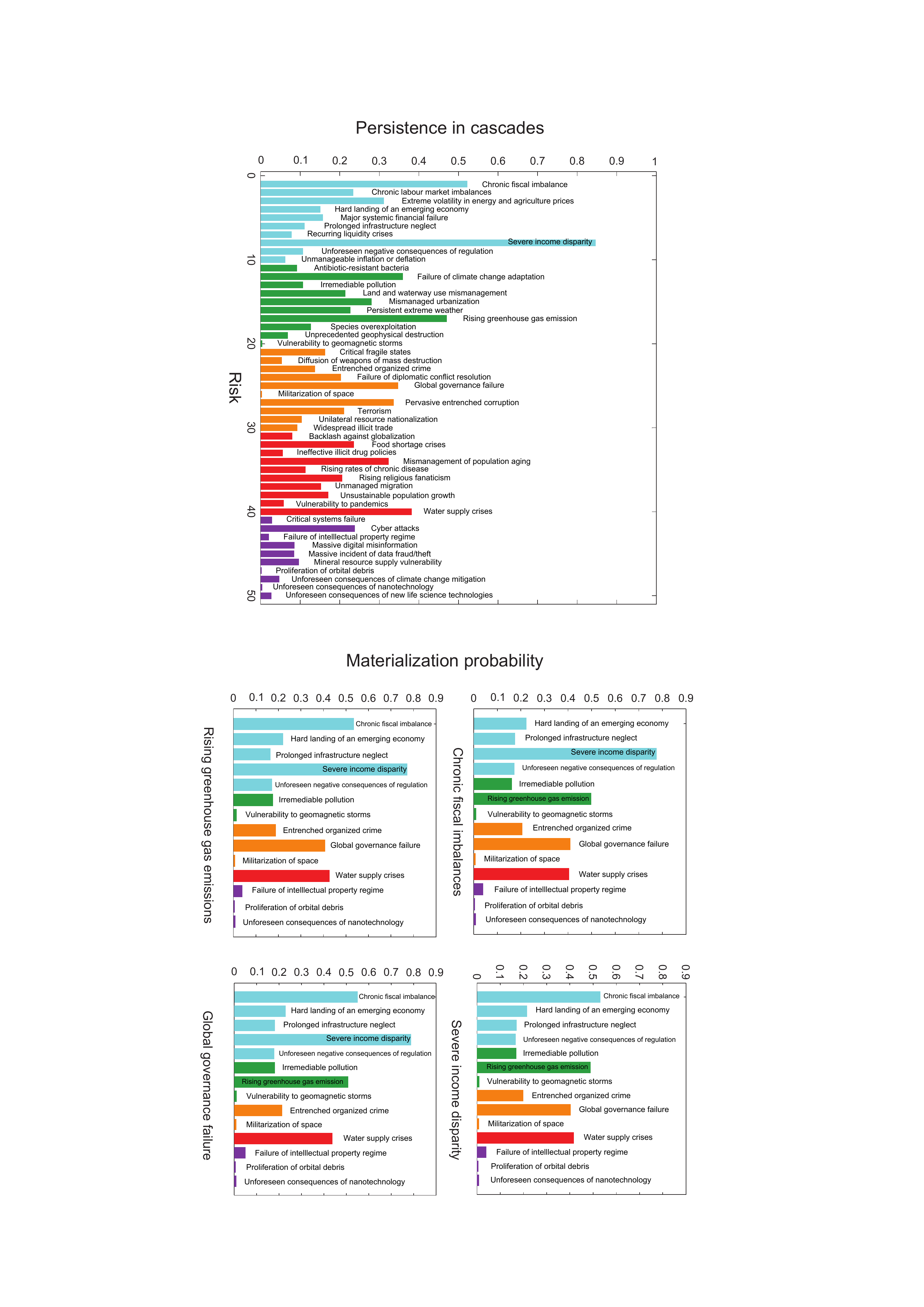}
\end{center}
\vspace{-0.75in}
\caption{{\bf Persistence and materialization probabilities of risks in cascades:} (a) The bar graph shows the fraction of the total lifetime of a cascade 
that a given risk is expected to be active, as obtained from $10^6$ simulations for each of $15$ different initiators, where initiators are chosen from 
sets of risks with high contagion potential, medium contagion potential and low contagion potential. The specific risks chosen as initiators were risks 
$1, 8, 9,12,16, 20, 23, 25, 26, 27, 31, 33, 42, 47, 49$. (b) The bar graphs show the materialization probabilities of four labeled risks, as a function 
of the initiator of the cascade (whose indices are shown above their respective bars). Each experiment ended when either the selected risk was infected, 
or all risks became inactive.
}
\label{persistenceplot}
\end{figure*}
%

\subsection*{Model dynamics}
In this section, we compare and contrast the dynamics of the network and disconnected models. In the later model, risks change their states only through 
internal Poisson processes of materialization and recovery. Thus, the dynamics of this model is simple, the activity level oscillates around the sum of 
average activity levels of risks. For risk $i$ this quantity, $a_i$, is defined as
\begin{equation}
a_i = \frac{p_i^{\rm int}}{p_i^{\rm con}+p_i^{\rm int}}
\label{persistent-disconnected}
\end{equation}
The resulting average activity level for the disconnected model is $10.75$. For comparison, the continuous time model is defined by the set of the following linear ODEs:
\begin{equation}
\frac{ds_i(t)}{dt}=\lambda^{\rm int}_i(1-s_i(t))-\lambda^{\rm rec}_is_i(t)
\label{linearODE}
\end{equation}
that has a simple solution: $\forall 1\leq i \leq N: s_i(t)=\frac{\lambda^{\rm int}_i}{\lambda^{\rm int}_i+\lambda^{\rm rec}_i}\left(1-e^{-(\lambda^{\rm int}_i+\lambda^{\rm rec}_i)t}\right)$.
The sum of these solution functions represents an average activity level of the model. It asymptotically tends to $\sum_{i=1}^N \frac{\lambda^{\rm int}_i}{\lambda^{\rm int}_i+\lambda^{\rm rec}_i}=10.66$, in good agreement with the discrete time Bernoulli process model.

In the network model activity level depends on the current state of the particular risks, as active risks influence inactive ones. If currently
active risks have low contagion potential, the system will drift to the lower activity level. Analyzing this case shows that this drift can cause all risks
to be inactive. On the other hand, active risks with high contagion potential will drift the system to even higher activity level than currently attained. 
Up to top $20$ risks with high contagion potential will cause the drift towards increasing activity level. Thus, the network model will
oscillate between the upper bound of $21$ active risks and a natural lower bound of having all risks inactive (albeit the latter is less likely than former, 
because highly contagious risks are also highly vulnerable to materialization via contagion). The average activity level for the network model is 
$13.79$, which is close to the value of $13.94$  yielded by the continuous time model defined by Eq.~\ref{ODE}.
The distribution of activity levels in both models with discrete time is close to normal, $N(13.79,3.71)$ 
for network model and $N(10.75,2.69)$ for disconnected model.  

\section*{Methods}

The first step to define the model is to relate the Poisson process intensities that determine the event probabilities in the model, 
to quantities provided by the expert assessments, namely, the likelihoods ($L_i$) of internal materializations 
of risks over a decade, and the influence ($b_{ij}$) that a given risk's materialization has on other risks. 

\subsection*{Mapping experts' assessments into Poisson process intensities}
We first normalize the likelihood values to probabilities in their natural range of $[0,1]$ by a simple linear transformation:

\begin{equation}
p_i = (L_i - 1)/4
\label{p10y}
\end{equation}

This normalized likelihood value $p_i$ is in direct proportion to the expert assessment $L_i$, and for our purposes captures the risk's vulnerability to 
failure. Next, we assume that the relationship between the probability, $p_i^{\rm int}$, that a risk $i$ materializes internally in a time unit (one 
calendar month) and this risk normalized likelihood value obeys the following polynomial form, with a parameter $\alpha$ defining the exact mapping:

\begin{equation}
p_i^{\rm int} = 1-(1-p_i)^{\alpha}
\label{pnot10y1}
\end{equation}

Thus, the probability of failing within a time period increases as the vulnerability $p_i$ increases, and Eq.~\ref{pnot10y1} coupled with our earlier 
assumption that the internal risk materialization is a Poisson process with intensity $\lambda_i^{\rm int}$, yields: 
\begin{equation}
\lambda_i^{\rm int} = -\alpha \ln(1-p_i)
\label{lambda-int}
\end{equation}

An advantage of Eqs.~\ref{pnot10y1},~\ref{lambda-int} is that their forms remain invariant under changes of the time-scale under 
consideration. Indeed, multiplying the original value of time unit by factor $f$ simply changes the Poisson process intensity and the value of $\alpha$ 
by the same factor $f$. For example, for the time unit of the expert materialization likelihood assessment set to a decade, the corresponding value 
of $\alpha$ is $120$ times larger than the value obtained when the time unit is set to a month. Moreover, the ratio of intensities is 
defined entirely by the ratio of the corresponding model parameters and is independent of the risk for which the corresponding 
Poisson processes generate events. So model parameters define the same ratio of intensities of all risks, while likelihood 
assessments define individual values of these intensities for each risk.  

Another advantage of the form of Eq.~\ref{pnot10y1} is that it can represent convex (for $\alpha >1$), linear (with $\alpha =1$), or concave (for $\alpha <1$) 
function, with the parameter $\alpha$ defining the shape that best matches a given set of historical data.

We adopt a similar reasoning to the mapping between the probability of continuation in a time unit $p_i^{\rm con}$ and 
the normalized likelihood values $p_i$. We start with the assumption that the probability of a materialized risk continuing over a time unit is 
$1-(1-p_i)^\gamma$, where parameter $\gamma$ defines the mapping from likelihood to probability. This dependence captures the increasing likelihood of 
continuation as the vulnerability $p_i$ increases and leads to the following equation:

\begin{equation}
\lambda_i^{\rm con} = -\gamma \ln (1-p_i)
\label{lambda-con}
\end{equation}

Finally, following similar arguments as above, the intensity $\lambda_{ji}^{\rm ext}$ of the Poisson process that enables a  materialized risk $j$ 
to influence the materialization of risk $i$ is a function of parameter $\beta$ defined as:

\begin{equation}
\lambda_{ji}^{\rm ext} = -\beta b_{ji}\ln(1-p_i)
\label{lambda-out}
\end{equation}

The factor $b_{ji}$ on the right hand side merely serves to capture the fact that the risks $i,j$ must be perceived by the experts as having an influence
on each other, in order for the probability of influence to be non-zero.

The forms provided in Eqs \ref{lambda-int}, ~\ref{lambda-con},~\ref{lambda-out} define the model completely, 
and all that remains is to fit the parameters $\alpha$, $\beta$, and $\gamma$ optimally to the historical data 
capturing the risk materialization events over the last $13$ years. In the historical dataset, each risk is assigned a state 
per month (the fundamental time unit) over the period of $2000-2012$. Thus, the likelihood of 
observing this particular sequence of risk materialization events through the dynamics generated by our model 
can be written as:
\begin{equation}
\mathcal{L}(\vec{S}(1),\vec{S}(2) \cdots, \vec{S}(T)) \equiv \prod\limits_{t=2}^{T}\prod\limits_{i=1}^{N} 
\mathcal{P}_i(t)^{S_i(t-1)\rightarrow S_i(t)}
\label{likelihood}
\end{equation}

where $T=156$ is the number of time units in the historical dataset and $N=50$ is the number of risks. Consequently, the logarithm of the likelihood 
of observing the sequence is:
\begin{equation}
\ln\mathcal{L}(\vec{S}(1), \vec{S}(2) \cdots, \vec{S}(T)) \equiv \sum\limits_{t=2}^{T}\sum\limits_{i=1}^{N}\ln ( \mathcal{P}_i(t)^{S_i(t-1)\rightarrow S_i(t)})
\label{log-likelihood}
\end{equation}

Following the well-known process of maximum likelihood estimation~\cite{Heidi,Dempster1977}, we find the arguments that maximize the 
log-likelihood to optimize the model fitness. For a given set of values of parameters $\alpha$, $\beta$, and $\gamma$, one can compute 
the log-likelihood of observing the given time-series of risk materialization using Eqs \ref{transitions} and ~\ref{log-likelihood}. 
Thus, by scanning different combinations of $\alpha$, $\beta$, and $\gamma$ over their respective feasible ranges, and by computing 
the resulting log-likelihoods, one can find with the desired precision the values of $\alpha$, $\beta$, and $\gamma$ that maximize 
the likelihood of observing the data. The likelihood function is smooth (see the plot in Figure~\ref{lsurface}) with a unique 
maximum that guarantees that found  parameter values are indeed globally optimal for the model considered. 
With the time unit of a decade, these optimal values (marked by $^*$ superscript) are $\alpha^* = 0.365 \approx 4/11, \beta^* = 0.14 \approx 1/7, 
\gamma^* = 427 $, and the log-likelihood of observing the data given these parameters is $-415.6$. We refer to so-defined model as {\it network model}.
\begin{figure*}[!htb]
\vspace{-3.75in}
\begin{center}
\includegraphics[width = 6in]{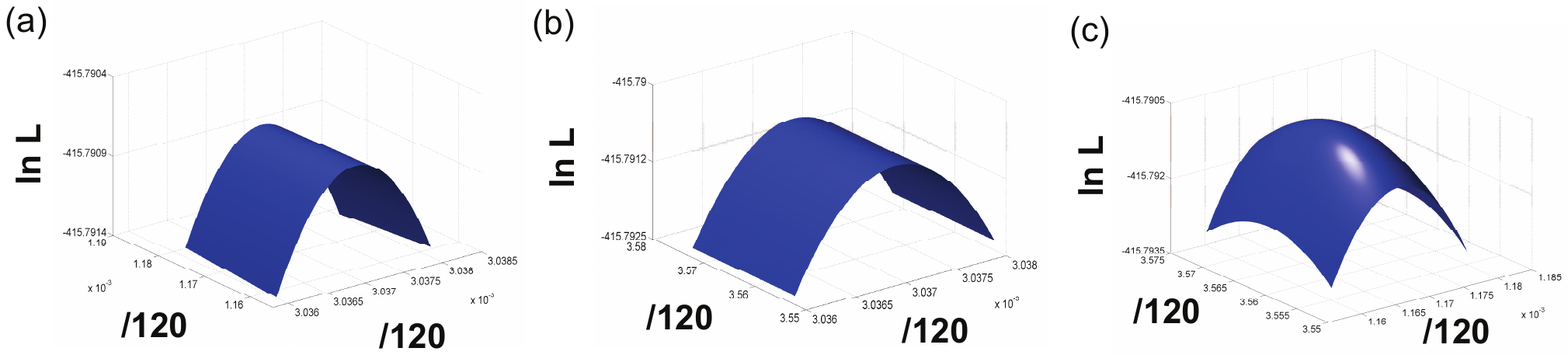}
\end{center}
\vspace{-3.75in}
\caption{{\bf Log-likelihood of data as a function of model parameters:}
Behavior of the log-likelihood of observing the historical sequence of risk materialization events as a function of pairs of model parameters. 
The three plots show the variation of $\ln \mathcal{L}$ for fixed $\beta$, $\gamma$ and $\alpha$ respectively as a function of the values of the 
two remaining parameters. The first plot has $\gamma$ fixed at its optimal value of $427$, the second plot has $\beta$ fixed at $0.14$ and the last plot has 
$\alpha$ fixed at $0.365$.
}
\label{lsurface}
\end{figure*}
%
%

\subsection*{Establishing model properties}
Next, we measure how vulnerable our model is to noise in the expert data. To do this, we randomly perturb each average likelihood value provided 
by the experts to a value within one standard deviation from the average, and create $10$ sets of such randomly perturbed likelihood data. Next, 
we compute $10$ parameter sets that maximize log-likelihood of observing the historical data. Then, we run $10$ models, termed noisy data models, 
defined by the obtained parameter sets. For each noisy model, we compute its monthly  activity level, which is the number of risks active in each 
month averaged over $10^6$ runs of this model. Finally,  we compute the maximum differences between parameters and monthly activity levels at each 
month of the network model and all $10$ noisy data models. The optimal parameters of noisy data models were within $\pm 1.4$\% of those values for 
network model. Average activity levels of these models were within $\pm 3$\% of the corresponding value for the network model. Finally, the maximum 
log-likelihoods of noisy data models are within $\pm 1$\% of this value for network model. These are 
small differences compared to the corresponding differences between the network and others models discussed below. 

\subsection*{Alternative models}
We also measure the importance of network effects by comparing the maximum log-likelihood obtained above to the corresponding maximum log-likelihood 
value for a model which is oblivious to network effects, so has $\beta = 0$. We refer to this model as the {\it disconnected model}. With the time unit 
of a decade (which experts used in their likelihood assessment), the two optimal parameters are $\alpha^d=0.91$, $\gamma^d=\gamma^*=427$ and maximum 
log-likelihood is $420.1$. Using  the likelihood ratio (LR) test~\cite{Pawitan}, we conclude the network model outperforms the disconnected model at 
a significance level of $0.01$. This result demonstrates that to fully uncover the value of expert data requires accounting for network effects, as 
we have done in our network model.
 
Setting $\alpha=1.0$ creates a model that we refer to as {\it expert data based model} which yields maximum log-likelihood of $420.1$ that is only 
slightly higher (by $0.02$\%) than for the optimal disconnected model. More importantly, it results in particularly simple form for one-decade risk 
$i$ materialization probability: $p_i^{\rm int} = p_i$. This linear mapping demonstrates that the averages of experts' assessments of risk 
materialization likelihoods are in fact excellent estimates of probabilities of risk failures in the ten-year period. It also attests to validity 
of our historical data and of expert assessments, since any mistake in those two datasets would make a mapping from expert data to probabilities 
a complex function. Similar high quality expert forecast in strategic intelligence was discussed in \cite{pnas2014}. Yet, this results uncovers the 
limit of expert assessments, as they capture the aggregate probability of failures resulting from internal and external risk materializations without 
ability to distinguish between them. Since external materialization depends on which risks are active, any change in the states of the risks changes 
such aggregate probability. Our model, through parameter mapping onto the historical data, is able to separate external and internal materialization 
probabilities and therefore is valid regardless of the current or future states of the risks. 

We also evaluate the value of experts' assessments of risks susceptibility to failures and their influence on each other for modeling risks. 
An alternative model with individual parameters for each risk susceptibility and influence would have too many degrees of freedom to be well-defined. 
However, the network model applied to risks with uniform likelihood and influence, a model to which we refer to as {\it uniform model}, and which is 
therefore agnostic to expert data, yields a maximum log-likelihood of $437.1$, far from  what the disconnected and expert data based models deliver. 
According to the LR test~\cite{Pawitan}, the disconnected and expert data based models cannot be distinguished from each other with any reasonable 
significance level. However, the same test allows us to conclude that these two models outperform a simple uniform model based only on historical 
data with a significance level of $0.001$.

Summary of models discussed here is provided in Table~\ref{model-table}.
\begin{table}
\begin{tabular}{ l | c | r }
\hline
Model & Parameters & Data used \\
  \hline                       
  network model & $\alpha, \beta, \gamma$ & $L_i, b_{ji}$, historical data\\
  disconnected model &  $\alpha, \gamma$ ($\beta = 0$) & $L_i$, historical data \\
  expert data based model & $\gamma$ ($\alpha = 1, \beta = 0$) & $L_i$, historical data \\
  uniform model & $\lambda^{\rm int}, \lambda^{\rm con}, \lambda^{\rm ext}$ & historical data \\
  \hline  
\end{tabular}
\caption{{\bf Summary of models: Parameters for the models mentioned in the text, and the data utilized to estimate the respective parameters.} 
Parameters $\alpha$, $\beta$, and $\gamma$ govern the Poisson process intensities for internal materializations, pairwise influence, and continuation, 
as expressed in Eqs.~\ref{lambda-int},~\ref{lambda-out}, and \ref{lambda-con} respectively. $L_i$ represents the likelihood score provided for risk 
$i$ by the WEF report, and $b_{ji}$ is a binary variable that adopts a value of $1$ if the materialization of risk $j$ is deemed to have an influence 
on the materialization of risk $i$ in at least one of the experts' opinion. The expert data based model is the disconnected model in which value of 
parameter $\alpha$ is restricted to $1.0$. The uniform model uses two parameters for the Poisson process intensities for internal materialization and 
materialization continuation (Eqs.~\ref{lambda-int},~\ref{lambda-con}) which are assumed to be identical for all risks. It uses the third parameter 
to define the influence probability between risks (Eq.~\ref{lambda-out}) which is assumed to be the same for all risk pairs. The network model 
outperforms all other models in explaining the observed data.}
\label{model-table}
\end{table}
 
\section*{Discussion}
To summarize, in this study we have presented a method of obtaining a quantitative picture of the global risk network, starting from the qualitative 
observations provided by $1000$ WEF experts. We assume a three parameter network model for the propagation of risk materialization (representing 
the corresponding network node failures), and obtain maximum likelihood values for the parameters using historical data on risk materialization.

Our model was built upon expert assessments available in the WEF report which enabled the construction of a detailed and heterogeneous weighted 
network of risks. As we show, ignoring network effects (i.e. the disconnected model) or ignoring specific heterogeneities in the failure likelihoods 
and influence (i.e. the uniform model) yielded poor results in comparison to the network model. This underscores the importance of the expert 
assessments in building a model capable of matching the available activity data well, and therefore yielding reliable insights. We have also found 
the greatest strength of expert assessments, which is nearly perfect forecast of aggregate failure probabilities of different risks, but also those 
assessments greatest weakness, which is inability to separate external risk materialization probabilities form internal ones. We have developed 
an approach in which by selecting proper model parameters and using maximum likelihood estimation to find optimal model parameters, we are able to 
do such separation.

We have uncovered the global risk network dynamics and measured its resilience, stability, and risks contagion potential, persistence, and roles 
in cascades of failures. We have identified risks most detrimental to system stability and measured the adverse effects of risk interdependence 
and the materialization of risks in the network. According to these studies, the most detrimental is risk $8$ -``Severe income disparity". Other 
risks that play a dominant role due to either their contagion potential or their persistent materialization are: $1$ - ``Chronic fiscal imbalances", 
$25$ - ``Global governance failure", $27$ -  ``Pervasive entrenched corruption", $12$ - ``Failure of climate change adaptation", $17$ - 
``Rising greenhouse gas emissions", and $40$ - ``Water supply crises".  

Utilizing the complete network model generated using the WEF data provides a much more detailed picture of the threat posed by different risks 
than the one obtained by simply relaying only on their failure-likelihood $L_i$ values and using the disconnected model. Additionally, our analysis 
demonstrates that the carrying capacity of the network i.e. the typical activity expected in the network given the current parameters, is about 
$13.7$ risks or $27\%$ of the total number of network nodes, of which four are persistently chosen from a subset of seven risks (see 
Figure~\ref{persistencenormal}). Aiming to reduce this overall carrying capacity could potentially be an overarching goal of global risk minimization. 

There are several prospects for extending the model that we presented here and its further analysis. First, obtaining more robust historical estimates 
of risk materialization may help us improve the fitting of the model. Secondly, it will be beneficial to account for slow evolution of network 
parameters in time. This change in network characterization will be captured by a model through expert data provided in yearly WEF reports, resulting 
in time dependent $L_i$s and $b_{ij}$s. Furthermore, the accuracy of the model could possibly be improved by assuming the existence of different 
dynamics for chronic risks as compared to sporadic risks. 

From a larger perspective, our attempt here has been to utilize data crowdsourced from experts towards gaining a quantitative picture of the network 
of global risks, which in turn has yielded some actionable insights. The network by definition has risks of varying complexity, which arguably makes 
the risk mitigation process more involved for some risks than for others. In such a scenario, our quantification of the relative impacts of different 
risks could provide a valuable guide to any cost-benefit analysis involved in the design of policies or strategies aimed at global risk minimization.

The ideal next step given the insights provided by our model would be for domain experts to provide tailor-made recommendations for the pertinent 
risks, such that the likelihood of systemic failures is strongly curbed. The effect of such recommendations can also be tested using our model. 
We therefore believe that the contribution of this paper is to implement the crucial step that lies between the gathering of crowdsourced data and 
the prescription of domain-specific recommendations.

\section*{Acknowledgments}
This work was partially supported by DTRA Award No. HDTRA1-09-1-0049, by the Army Research Laboratory under Cooperative Agreement Number W911NF-09-2-0053, and by the EU's 7FP under grant agreement no 316097. The views and conclusions contained in this document are those of the authors and should not be interpreted as representing the official policies either expressed or implied of the Army Research Laboratory or the U.S. Government. 

\section*{Author contributions}
Designed research: BKS, XL;
Performed research: BKS, XL, AA, SS;
Analyzed data: BKS, XL;
Designed simulation software: BKS, XL;
Wrote the paper: BKS, SS.

\section*{Competing Financial Interests Statement}
The authors declare no competing financial interests.

}

\end{document}